# Machine Learning for Location Prediction Using RSSI On Wi-Fi 2.4 *GHz* Frequency Band Technical Report

Ali Abdullah S. AlQahtani
*Computer Systems Technology*
North Carolina A&T State University
Greensboro, North Carolina, USA
AlQahtani.aasa@gmail.com

Nazim Choudhury
*Computer Science*
University of Wisconsin-Geen Bay
Green Bay, Wisconsin, USA
Choudhun@uwgb.edu

*Abstract*—For decades, the determination of an object's location has been implemented utilizing different technologies. Despite GPS (Global Positioning System) provides a scalable, efficient and cost effective location services however, the satellite emitted signals cannot be exploited indoor to effectively determine the location. In contrast to GPS which is a cost-effective localization technology for outdoor locations, several technologies have been studied for indoor localization. These include Wireless Fidelity (Wi-Fi), Bluetooth Low Energy (BLE), and Received Signal Strength Indicator (RSSI) etc. This paper presents an enhanced method of using RSSI as a mean to determine an object's location by applying some Machine Learning (ML) concepts.The binary classification is defined by considering the adjacency of the coordinates denoting object's locations. The proposed features were tested empirically via multiple classifiers that achieved a maximum of 96% accuracy.

*Index Terms*—access point, RSSI, indoor localization, machine learning, supervised classification, dynamic time warping.

## I. INTRODUCTION

LOCALIZATION has attracted a lot of research effort in the last decade due to its pertinence to different sorts of commercial applications. The surge in academic research and industrial applications of localization is not only due to the explosion of location based services but also for the location information which is typically found useful for coverage, deployment, routing, target tracking and rescue. An object's position can be identified by utilizing different technologies. Some of the technologies can be used only to determine outdoor positions, such as Global Positioning System (GPS), while some can be used to determine both indoor and outdoor positions. The latter technologies include Bluetooth, Wireless Fidelity (Wi-Fi), and RSSI (Received Signal Strength Indicator) etc.

*A. GPS*

GPS is an outdoor navigation system that depends on thirty satellites orbiting the Earth. GPS was developed for military purposes. However, nowadays, anyone can use it as long as he/she owns a GPS-enabled device. Furthermore, at least three GPS satellites have to be available to determine an object's position. One advantage of GPS is that it provides a location with high accuracy (i.e., up to five meters precisely); it also operates with no specific infrastructure required. The downside of GPS is that it consumes quite some power due to the long-distance communication with satellites around the Earth. In addition, weather situations can affect the communication.

*B. Bluetooth*

Bluetooth is a wireless communications technology that works in short-range. As a wireless short-range communications technology, the devices must be within approximately ten meters of each other. For two decades, Bluetooth has been around and its newest version is BLE, which is making significant strides in positioning and most of today's smartphones support. The advantages of BLE are that interaction is not required, it works accurately outdoors and indoors, and it requires low energy to function. However, one of BLE's disadvantages is that it is a little less accurate than GPS.

*C. Wi-Fi*

Wi-Fi positioning systems are one of the geolocation systems that depend on nearby Wi-Fi access points to determine where an object is located as [1]. Usually, Wi-Fi-enabled devices are connected to access points through a specific radio frequency, usually 2.4GHz or 5.0GHz. The range of these radio waves can be up to one hundred meters, which means Wi-Fi can be used on both indoor and outdoor positioning systems. The advantages of using Wi-Fi are that it consumes low energy, is accurate up to ten meters (depending on the availability of Wi-Fi networks), and requires no additional infrastructure. However, one of its disadvantages is that it might require paid service.

In short, the discussed techniques are compared, see TABLE I. Throughout the comparison, GPS are: (1) An accurate outdoor positioning (up to five meters); (2)It works everywhere outdoors (3) It does not require infrastructure; (4) It requires High energy; (5) It Often interrupts by weather-related situations; (6) It does not work indoors. Bluetooth are: (1) An accurate indoor/outdoor (depending on infrastructure); (2) It requires Low energy (1/15th of GPS); (3) It requires Minimal wireless infrastructure; (4) access to Bluetooth is required. Wi-Fi are: (1) Accuracy depending on the availability of Wi-Fi; (2) Low energy use (1/10th of GPS); (3) it might require paid service or a known local infrastructure network.

## II. LITERARY REVIEW

Several methodologies and solutions have been proposed to determine location. Each methodology and solution are unique, solving the problems in their own unique way.

As of today, there are numerous numbers of research that utilize RSSI to determine an objects' position. In 2021, two papers were published utilized beacon frames to prove the location of users [2] [3] [4]. Both of them present a method that scans for the RSSI value then plug it into mathematical equation (i.e., Trilateration System)to determine a user's position.

The authors present an RSSI-based Distributed Real-time Indoor Positioning Framework [5]. The Path Loss Long Equation E.q.1 in the proposed framework is modified to transfer RSSI value to a distance to form a cost-effective solution in BLE environment.

$$PL_{log} = PL_0 + 10\gamma \: log_{10} \frac{d}{d_0} \quad (1)$$

In 2019, Wang, et al. published a paper titled *"Tun-nel Vehicle RSSI Positioning Algorithm Based on LMLF Model"* [6]. The Presented scheme utilizes LMLF (Linear Modified Log Function) algorithm to improve objects' position accuracy depending on RSSI to provide a safety solution while driving in tunnels.

Nakai, Kawahama, and Katsuma in 2018 presented a method to reduce positioning errors in an environment with four or more nodes [7]. Utilizing unstable RSSI of short-range communication, The proposed approach reduces positioning errors.

In 2020, a paper titled *"Improved RSSI based Angle Localization using Rotational Object"* was published [8]. The authors depend on a rotating object (e.g., a rotating ceiling fan) to find angles in the indoor environment with the most accurate and reliable RSSI values.

RSSI technology has been studied to determine personnel localization in metal underground mines [9]. The discussed approach obtains the distribution of underground personnel in real-time for production safety in the underground mine. Moreover, several papers were published that depend on RSSI technology to implement secure authentication systems [1], [10]–[13].

In next section, the presented scheme backgrounds are discussed; each one in its own subsection.

## III. EXPERIMENT ENVIRONMENT

### A. Problem formulation

In this study, we address the location determination problem as a binary classification task. A pair of objects/individuals can belong to either of the two classes: (i) adjacent and (ii) distant. Both objects/individuals are considered adjacent to each other if they belong to the same room. On the contrary, the pair will be considered as distant if they are located in different rooms. Location-based feature vectors were constructed by considering the received signal strengths (RSSI) of the broadcast signal from three access points. Feature vectors were input into the binary supervised classifiers to classify the pair's proximity and thus predict the location of the pairs.

### B. Data collection

The experiment was conducted in two adjacent rooms with an academic building as displayed in Figure 1; the red lines present both axis. The coordinates $(-x, y)$ and $(x, y)$ denotes different locations of the mobile devices in the left and right room respectively. The blue diamonds represent the access points. the blue diamonds present the placements of three access points. The first room (right side of the vertical axis in the figure) with both positive horizontal and vertical coordinates had dimension of 35 feet by 25 feet. The second room (left side of the vertical axis in the figure) had the dimension of 33 feet by 25 feet. The experiment included the followings: The proposed scheme architecture consists of:

1) Participating devices: Wi-Fi-enabled devices (e.g., a smartphone with a relevant application installed on it or a Raspberry Pi).
2) Wireless access points: Three access points (APs) are located in three different locations within the first room. Access points were utilized to determine the positions of the participating devices within a room. Participating devices recorded the RSSI values which were automatically sent off to a remote server.
3) A remote server collected the RSSI readings sent by the Wi-Fi-enabled devices.

TABLE I: Comparison Technologies.

|  | GPS | Bluetooth | Wi-Fi |
|---|---|---|---|
| Outdoor & Indoor Positioning | Outdoor Only | Outdoor & Indoor | Outdoor & Indoor |
| Energy Consumption | High | Medium | Low |
| Positioning Accuracy | Up to $5m$ | $< 25m$ | $< 25m$ |
| Required Infrastructure | No | No | Minimal |
| Ideal Environment | Outdoor Everywhere | Outdoor\|Indoor Uncontrolled | Outdoor\|Indoor Controlled |

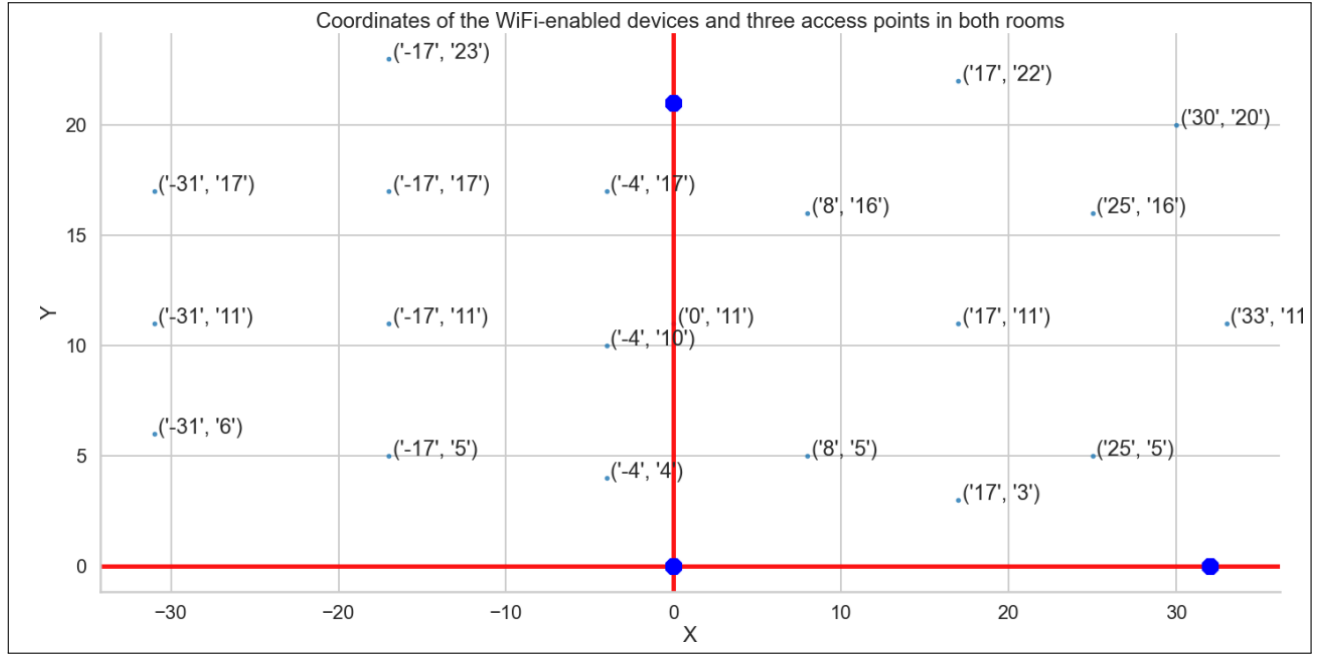

Fig. 1: Coordinates of the Wi-Fi-enabled devices in two adjacent rooms.

The coordinates in the figure are the Wi-Fi-enabled devices in each room. Each room had 10 devices. The access point were placed at the locations (0,0), (0,21) and (32,0), close to the walls. On each point within each room, we scanned the RSSI from each access point. Multiple trials (i.e., 10) were taken to collect the RSSI data and each trial was collecting data at every four seconds interval. The distributions of all unique RSSI values, collected from each access points against each point that worked as the building blocks of our machine learning features/input variables were saved in a server. In Figure 2 presents the unique measurements of RSSI signals for two points (-17, 11) and (-31, 11) collected via three access points (i.e., $AP_1$, $AP_2$ and $AP_3$).

### C. Supervised learning

In this article, we defined the supervised binary classification by considering the adjacency of the coordinates. Two coordinates belonging to the same room are adjacent whereas two coordinates belonging to the different rooms are distant from each other. Therefore, two classes (response variable) in this binary classification problem is "adjacent" and "not adjacent/distant". The input variables were computed via feature engineering process over the collected RSSI values.

### D. Feature engineering

We applied different methods to transform the raw distribution of RSSI values into meaningful features that capture well the inherent structures of the underlying problem and to improve the classification performance.

*1) Statistical Methods:* We used the average (mean), mode, minimum value of the absolute value of the unique RSSI values (measured in nagative numbers) aggregated for two coordinates. We also considered the difference of the mean (average) values of unique and absolute RSSI values collected for every individual coordinate.

*2) RSSI strength-based Methods:* RSSI measurements represent the relative quality of a received signal on a

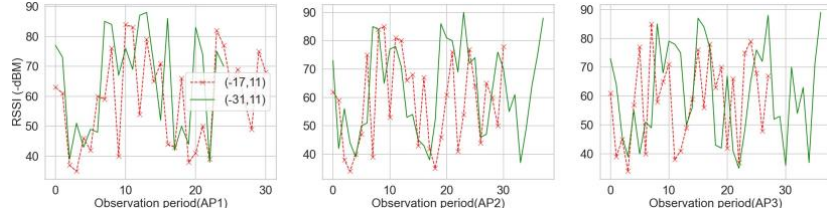

Fig. 2: Unique RSSI signals at two different points.

device. The higher the RSSI value, the stronger the signal. When measured in negative numbers, the number that is closer to zero usually means better signal. As an example -50 is a pretty good signal, -75 is fairly reasonable, and -100 is no signal at all. By considering this, we computed the relative strength of signal per coordinate using the distribution of the RSSI values collected for that corresponding coordinate. We computed the rate of RSSI values less than or equal to -50 dBm compared to the toal number of RSSI values to measure the higher level of signal strength. We also considered the rate of RSSI values less than or equal to -50 dBm compared to the toal number of RSSI values to measure the average level of signal strength.

*3) Dynamic time warping (DTW):* In time series analyses, DTW provides intuitive distance measurements between temporal sequences by ignoring both global and local deviations in the time dimension [14]. It is a well-known technique to find an optimal alignment between two given (time-dependent) sequences where the sequences are warped intuitively in a nonlinear fashion to match each other. It was found successful to automatically cope with time deformations and different speeds associated with time-dependent data when applied in studies in data mining and information retrieval [15]. DTW measures the similarity between two time series by shrinking or expanding, or simply warping the time axis of one (or both) sequences to achieve better alignment. In this study, these two time series denote the sequences of RSSI values collected by the Wi-Fi-enabled devices at two different points (see Figure 1 for each access point. The optimal warping path between these two time series is defined as a warping path with the minimum distance among all possible warping paths. To accomplish that it may encounter that a single point in one time series may be mapped to multiple points of the other. It is noteworthy that DTW can measure the similarity between two time series of dissimilar length. In Figure 3, this study presents a comparable representation of calculating similarities between two temporal sequences using traditional distance measures (Figure 3a) (e.g., Euclidean) and DTW method (Figure 3b). In Figure 3, dashed lines represent the distance between corresponding points in both time series.

### E. List of Features

By considering two sequences of unique RSSI values $X_i$ and $Y_i$ belonging to two Wi-Fi-enabled devices $X$ and $Y$, we computed the features described in Table II. Each device is denoted by a point coordinates $(x, y)$. It is noteworthy that all these features are computed by considering the RSSI values collected from three access points. Thus, for three access points we have 18 features altogether for each pair of coordinates.

### F. Performance metrics

We relied on the confusion matrix which is one of the most intuitive and easiest metrics used for finding the correctness and accuracy and summarizing the performance of the classification model. A confusion matrix is an N X N matrix, where N is the number of classes being predicted. The matrix compares the actual target values with those predicted by the machine learning model. This gives us a holistic view of how well our classification model is performing and what kinds of errors it is making. For binary classification problem, we would have a 2 x 2 matrix Figure as shown in 4. True Negative which shows the number of negative examples classified accurately. Similarly, True Positive which indicates the number of positive examples classified accurately. False Positive represents the number of actual negative examples classified as positive; and False Negative value which is the number of actual positive examples classified as negative.

The confusion matrix visualizes the relations between the classifier outputs and the true ones. Using this confusion matrix, we computed the following metrics to determine the performance of our classification models

1) Accuracy: One of the most commonly used metrics while performing classification is accuracy. The accuracy of a model (through a confusion matrix) is calculated as below: $Accuracy = \frac{TP+TN}{TP+TN+FP+FN}$
2) F1 score: F1 Score tries to find the balance between precision and recall. Precision attempts to answer what proportion of positive identifications was actually correct. Recall (True Positive Rate) attempts to answer what proportion of actual positives was identified correctly F1 score denotes how precise the classifier is (how many instances it classifies correctly), as well as how robust it is (it does not miss a significant number of instances). The greater the F1 Score, the better is the performance of our model. Mathemically, F1 score is $F1 = 2 * \frac{1}{\frac{1}{Precision} + \frac{1}{Recall}}$

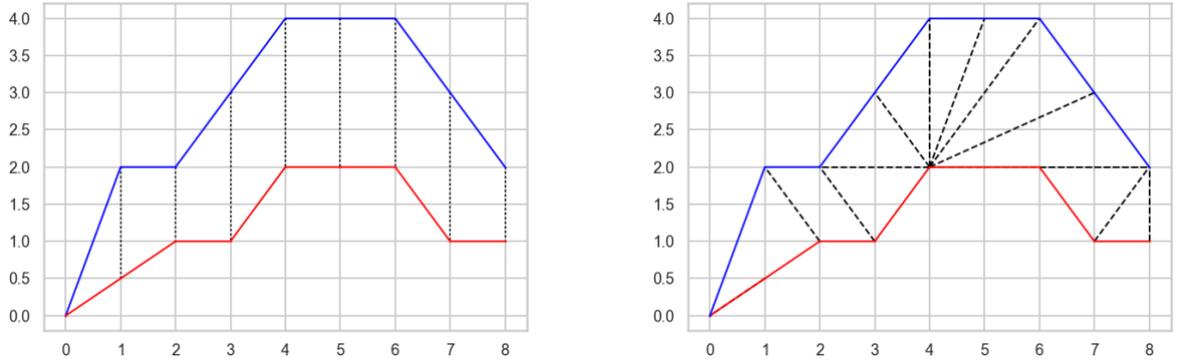

(a) Traditional Euclidean  (b) Dynamic Time Warping

Fig. 3: Visualizations of measuring time series similarity between two temporal sequences.

TABLE II: Names and descriptions of the features computed in this study

| Feature Name | Feature Description |
|---|---|
| Difference in mean $MD(XY)$ | The difference between the average of the RSSI values acquired by the devices $X$ and $Y$. $\|X - Y\|$ where $X = \sum_{i=0}^{n} \frac{X_i}{n}$ |
| Mean signal strength $S^{avg}$ | The average of the unique RSSI values aggregated for both the devices $X \cup Y$ |
| Minimum signal strength $S^{min}$ | The minimum of the unique RSSI values aggregated for both the devices $min(X \cup Y)$ |
| High signal strength $RSSI^{high}$ | The ratio between the number of RSSI values less than or equal to -50dBm and all RSSI values acquired by both devices. $RSSI^{high} = \|\{s : s \in X \cup Y, s <= -50dBm\}\| / \|X \cup Y\|$ |
| Average signal strength $RSSI^{avg}$ | The ratio between the number of RSSI values less than or equal to -70dBm and all RSSI values acquired by both devices. $RSSI^{avg} = \|\{s : s \in X \cup Y, s <= -70dBm\}\| / \|X \cup Y\|$ |
| Signal similarity $DTW(XY)$ | The similarity between two sequences of RSSI values $X_i$ and $Y_i$ belonging to two coordinates denoting the locations of two devices measured by the dynamic time warping method |

Fig. 4: Confusion matrix

### G. Classification dataset and classifiers

We constructed a classification dataset of 300 samples where 100 samples were positive (1) and 200 samples were negative (0). Here the positive class denotes the adjacency between two points and the negative class denotes the non-adjacency between two points. We used five different classifiers, namely, logistic regression (LR), K-Nearest Neighbors (KNN), Random Forest (RF), Support Vector Machines (SVM) with radial basis function (RBF) kernel and Decision Trees (DT). In decision tree, we used the Gini impurity function to measure the quality of a split. We left the maximum depth to the default which allows the nodes to be expanded until all leaves are pure or until all leaves contain less than minimum samples to split samples which was set to two. The DT classifier was configured to consider all the available features when looking for the best split. The feature values were normalized with zero mean and unit variance. The ratio between training and test samples was set to 75/25. This means 75% of the total samples were used as training samples to train the classifiers. The rest 25% was set aside as test samples to evaluate the performance of the trained classifiers. We used 10-fold cross-validation which is a technique to avoid overfitting and evaluate predictive models by partitioning the original samples into a training set to train the model, and a test set to evaluate it.

### IV. RESULTS

#### A. Classification performance

In Table III, we present the performance demonstrated by five different classifiers considering different performance metrics for both positive and negative classes. The reason behind the performance difference demonstrated by different classifiers is out of the scope of this study. However, from the table, it is observable that both RF and DTs demonstrated the best performance and exceeded other classifiers. In Figure 5, we present the summary of prediction results demonstrated by the five different classifiers in predicting both classes of samples. It is also observable that all classifiers were successful to classify the negative samples accurately (F1 score). However, the linear classification algorithm logistic regression and KNN under-performed in successfully predicting the positive samples' class label.

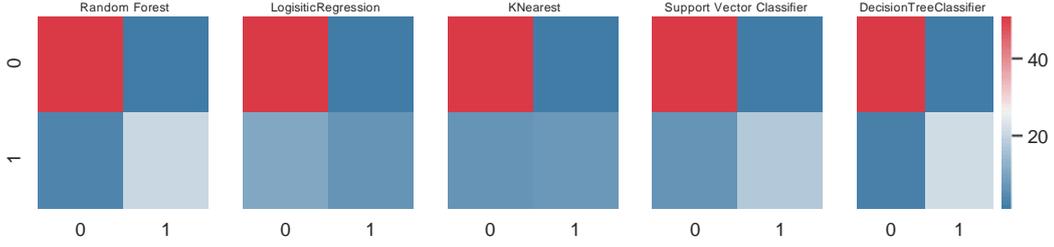

Fig. 5: Confusion matrix to summarize the performance of five classifiers used in this study.

TABLE III: Performance of different classifiers

| Metrics | class | LR | KNN | RF | SVM | DT |
|---|---|---|---|---|---|---|
| Accuracy | n/a | 76% | 75% | 96% | 90.6% | 94.6% |
| F1 Score | 0 | 0.84 | 0.83 | 0.97 | 0.94 | 0.96 |
| | 1 | 0.53 | 0.57 | 0.93 | 0.83 | 0.91 |

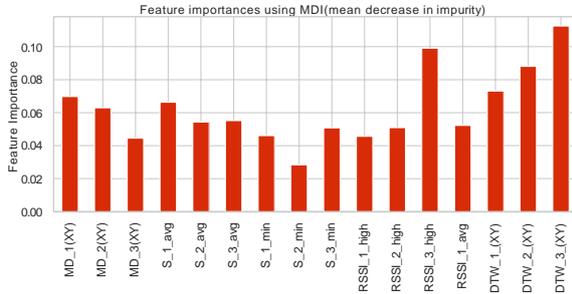

Fig. 6: Impurity-based importance of different features in classifying positive and negative samples by RF classifier

### B. Feature importance

Considering the performance demonstrated by the RF classifier exceeding other classifiers, we were also interested to identify the feature (variable) importance. Feature importance refers to techniques that assign a score to input features and describes which features are relevant based on how useful they are at predicting a target variable by the RF classifier. The RF algorithm has built-in feature importance which can be computed in two ways: (i) impurity importance and (ii) permutation importance [16]. In this study, we will consider only the impurity importance. The impurity importance is also known as the mean decrease of impurity (MDI). RF is a set of DTs where each DT is a set of internal nodes and leaves. This could be be the underlying reason behind the improved performances demonstrated by both RF and DT classifiers.

Gini index is commonly used as the splitting criterion in classification trees. In the internal node, the selected feature is used to decide how to divide the data set into two separate sets with similar responses within. The features for internal nodes are selected based on Gini impurity or information gain. We can measure how each feature decrease the impurity of the split (the feature with highest decrease is selected for internal node). For each feature we can collect how on average it decreases the impurity. The average over all trees in the forest is the measure of the feature importance. In Figure 6, we present the feature importance by considering the mean decrease in impurity (Gini importance) of each feature used in the RF classifier. The indices associated with each feature (i.e., 1,2,3) denotes the three access points from which the RSSI values were emitted. From this figure, it is evident that DTW-based features and RSSI signal strength-based features performed well. Note that the features constructed from RSSI signals acquired from the third access point demonstrated better performance and exceeded the other features. Exploring the reason behind this performance improvement by the features computed from the RSSI values emitted by different access points is out of the scope of this study. Future study can explore the importance and signal strength of the access points in feature construction.

### C. Feature Distribution

In Figure 7, we present the Kernel Density Estimate (KDE) plots to visualize the distribution of feature values in the classification dataset. Figure 7 shows KDE plot to demonstrate the distribution of feature values of both positive (1) and negative (0) samples. Only high performing features computed by the dynamic time warping method and relative strength of the RSSI signal were considered. KDE presents a different solution to the same problem that is addressed by histograms. Rather than using discrete bins in histograms, a KDE plot smooths the observations with a Gaussian kernel, producing a continuous density estimate. We only considered the highest performing features described in the aforementioned section. These are mainly the temporal similarity-based features computed via dynamic time warping method and the ratio of high strength RSSI signals vs all signal values emitted by the third access point ($AP_3$). Classifiers can pick patterns where there is significant difference between both class distributions. The overlapping region between the class density of the positive and negative samples in these figures were the reason for the mis-classification. In all cases, we can observe that the overlapping regions were comparably lower contributing towards the higher classification performance demonstrated by these features.

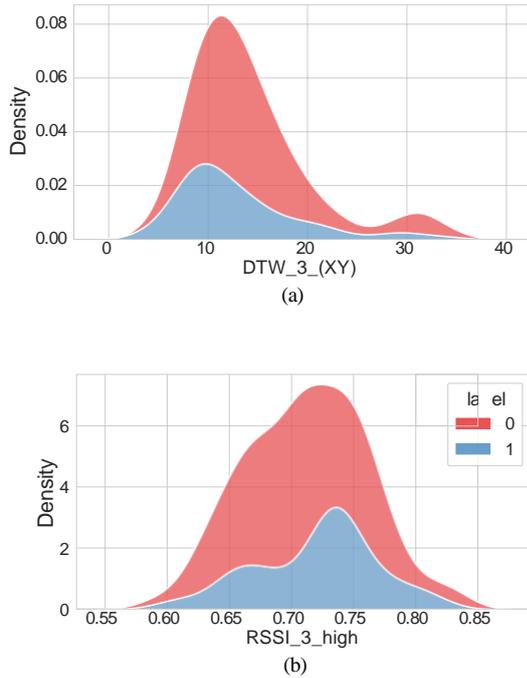

Fig. 7: KDE distribution of feature values

## V. CONCLUSION

Determining the location of either stationary or mobile objects as accurately as possible has already been a problem of interest for the scholars. The ability to predict users' location proximity plays a vital role in Location-Based Services and recommender systems. In this study, we trained different binary supervised classifiers to develop classification models for predicting location proximity of individuals/objects. Our models were applied on the test data and demonstrated a performance improvement of 96% (RF and DT classifiers.) We also identified the feature importance to determine which features contributed to this performance improvement. It turned out that DTW and RSSI signal strength based features were non-trivial in predicting the position of a device. Despite great performance scores, this study is not free from its limitation. We only considered small number of devices in both rooms (10 devices) which is due to the smaller class size. We used three access points to emulate the tri-lateration approach. Future studies can explore the impact of more access points and mode devices in the classification/prediction performance. Also, why features from different access points contribute differently in the performance can be a great research question for further exploration. We considered the marks of -50dBM and -70dBm as the measurement of good and avergae signal strength, however, future studies can consider other signal values as threshold for the same purpose and explore the performance of the prediction.